\documentclass[aps,prb]{revtex4}

\begin{document}

\title{D-wave superconductivity in doped Mott insulators}
\author{L. B. Ioffe and A. J. Millis}

\date{Dec 21, 2001}

\address{Center for Materials Theory \\
Department of Physics and Astronomy\\
Rutgers University\\
136 Frelinghuysen Rd, Piscataway NJ 08854}

\begin{abstract}
The effect of proximity to a Mott insulating phase on the charge transport
properties of a superconductor is determined. \ An action describing the low
energy physics is formulated and different scenarios for the approach to the
Mott phase are distinguished by different variation with doping of the
parameters in the action. A crucial issue is found to be the doping
dependence of the \textit{quasiparticle charge} which is defined here and
which controls the temperature and field dependence of the electromagnetic
response functions. Presently available data on high-T$_{c}$ superconductors
are analysed. The data, while neither complete nor entirely consistent,
suggest that neither the quasiparticle velocity nor the quasiparticle charge
vanish as the Mott phase is approached, in contradiction to\ the predictions
of several widely studied theories of lightly doped Mott insulators.
Implications of the results for the structure of vortices in high-T$_{c}$
superconductors are determined.
\end{abstract}

\maketitle

\section{\protect\bigskip Introduction}

High-$T_{c}$ superconductors are created by doping an antiferromagnetic
'Mott insulating' parent material \cite{Orenstein00}. A Mott insulator is a
material which band theory would predict to be metallic but which, because
of electron-electron interactions, is insulating \cite{Imada98}. It is
conventional to distinguish the different doping levels as 'overdoped',
'optimally doped', and 'underdoped'. Optimal doping is generally defined as
the carrier concentration which maximizes the resistively-defined
superconducting transition temperature $T_{c}$. Overdoped materials have
more added carriers and underdoped materials have fewer. All high-$T_{c}$
materials exhibit behavior which deviates from the
'Fermi-liquid-Migdal-Eliashberg' theory which describes most metals;
however, the deviations grow more pronounced as the doping is reduced
towards the Mott insulator, and indeed understanding the physics of the
underdoped regime (in other words, of the approach to the Mott insulating
phase) has emerged as one of the crucial questions in the high-$T_{c}$ field.

High-$T_{c}$ superconductivity remains a controversial subject \cite
{Orenstein00}. However, there is general agreement that one important phase,
which may actually be observed in sufficiently clean materials, is a
homogeneous superconducting phase characterized by an energy gap vanishing
only along the nodal directions ($p_{x}=\pm p_{y}$ in a material with
tetragonal symmetry) and possessing conventional quasiparticle and
supercurrent excitations. The low temperature behavior of a d-wave
superconductor is described by a general action, which depends on four
parameters which are defined below. The behavior of the parameters as the
Mott phase is approached is seen to reveal information about the underlying
physics of the Mott transition and governs the structure of vortices and
thereby the doping dependence of such macroscopic quantities as the upper
critical field. In this paper (which is largely a review) we describe the
information which may in principle be obtained from the low T properties,
summarize the present status of the experimental values of the different
parameters, and (although the presently available data are neither fully
consistent nor complete) outline the apparent physical implications of the
results. Two subsequent papers are planned: one presenting the derivation of
the general low temperature action from different microscopic theories and
one using it to analyse the vortex properties (in particular $H_{c2}(T)$ and
paraconductivity) in more detail.

Specifically, in this paper we consider a homogeneous d-wave
superconducting phase, assuming in particular that there is no spontaneously
broken time reversal symmetry and \ there are no other relevant excitations
apart from the quasiparticle and phase fluctuations. (We also restrict
attention to two dimensional models, but this restriction can easily be
lifted if desired). We find that a crucial parameter is what we define as
the effective charge of the quasiparticles. This can be determined at low $T$
by relating the observed electronic specific heat and photoemission (which
essentially measure the number and velocity of quasiparticles) to the
temperature dependence of the London penetration depth (which is related to
the ability of these states to carry an electrical current). Different
scenarios for the approach to the Mott transition produce striking
differences in the variation, with doping, of this charge. In particular,
theories involving some form of spin charge separation seem to lead to a
vanishing of the quasiparticle charge as the Mott phase is approached. We
analyze available data to determine which scenario actually occurs.

One very important feature of a superconductor is the structure of vortices
introduced by thermal fluctuations or via a magnetic field. A vortex is
characterized by a core, which may be defined in two ways: either via the
density of quasiparticle states, which is higher near the core and drops as
one moves away from the core region, or via the supercurrent, which varies
as $1/r$ far from the core and drops to zero inside the core. In several of
the theoretical approaches to the physics of the lightly doped Mott
insulator (including the one which seems to best fit the data discussed
below), the core size \textit{as defined from the supercurrent} may become
very large. This, combined with the behavior of the quasiparticle charge,
has remarkable implications for the size of the critical region, for the
behavior of the upper critical field $H_{c2}$ and for the physics of the
superconductor-insulator transition which must occur as the doping goes to
zero. These implications were pointed out in \cite{Lee96} and will be
further analysed by us in a subsequent paper.

The rest of this paper is organized as follows. In section II we present the
low energy, long wavelength theory of a d-wave superconductor near a Mott
insulator. In Section III we discuss the available experimental evidence
concerning the value and doping dependence of the parameters of the theory.
Section IV presents the limits of validity of the low energy action we
discuss, along with discussion of the physics occurring when these limits
are exceeded. A conclusion discusses the physical implications of our
formalism and findings. The Appendix gives the details of calculations of
the field dependence of the specific heat and superfluid stiffness in the
vortex state.

\section{Low Energy Theory}

At low temperatures the state of all (even underdoped)\ superconducting
cuprates seems to be a conventional $d_{x^{2}-y^{2}}$ superconductor \cite
{Tseui98,dinunderdoped,Shen95}: it is described by the usual low energy
degrees of freedom, namely a superconducting phase variable $\phi (r,t)$
and, as will be discussed in more detail below, apparently conventional
fermionic quasiparticle excitations\cite{Taillefer}. Gradients of the phase
correspond to supercurrents. Longitudinal supercurrents lead to charge
fluctuations which are coupled by the long-ranged Coulomb interaction.
Transverse supercurrents are described by a \textit{phase stiffness} $\rho
_{s}$ whose long-wavelength limit may be deduced from measurements of the
London penetration depth. In high-$T_{c}$ materials, the low-$T$ limit of $%
\rho _{S}$ is strongly $x$-dependent, vanishing roughly linearly as $%
x\rightarrow 0$. This behavior is understood as a consequence of the
suppression of charge fluctuations as the Mott insulating phase is
approached, and appears to be related to the decrease of the resistively
defined $T_{c}$ as $x\rightarrow 0$. Indeed in the very early days of high-$%
T_{c}$ Uemura and co-workers showed that in underdoped materials the ratio $%
T_{c}(x)/\rho _{s}(T\rightarrow 0,x)$ was essentially $x$-independent \cite
{Uemura89}. At roughly the same time this behavior was shown by a number of
workers to follow naturally from theoretical models of superconductivity
near a Mott transition \cite{Kotliar}, and later Emery and Kivelson argued
that the behavior could be understood in a more model-independent way as a
consequence of the unusually small phase stiffness characteristic of high-$%
T_{c}$ materials \cite{Emery95}. Recently \ Corson et. al reported direct
evidence that in underdoped $Bi_{2}Sr_{2}Cu_{2}O_{8}$ the superconducting
transition was indeed of the vortex-unbinding type driven by a small phase
stiffness \cite{Corson98}.

At scales less than maximal value of the gap, $\Delta _{0}$, the physics of
a two dimensional superconductor with tetragonal symmetry and a $%
d_{x^{2}-y^{2}}$ gap function is described by an effective Lagrangian
density $\mathcal{L}$ involving the phase $\phi $ of the superconductor and
quasiparticles excited out of the superconducting condensate: 
\begin{equation}
\mathcal{L=L}_{\phi }+\mathcal{L}_{F}+\mathcal{L}_{mix}  \label{L}
\end{equation}
Here $\mathcal{L}_{F}=\partial _{\tau }-H_{D}$ is the usual Dirac action
describing the 'nodal quasiparticles' excited in the vicinity of the nodes
of the d-wave gap function. \ In a superconductor, the fermionic energy
spectrum is given by $E_{p}=\pm \sqrt{\varepsilon _{p}^{2}+\Delta _{p}^{2}}$
with $\varepsilon _{p}$ the energy spectrum of the underlying fermions and $%
\Delta _{p}$ the gap function. For a $d_{x^{2}-y^{2}}$ gap function, $\Delta
_{p}$ vanishes linearly in the four nodes, i.e. for $\overrightarrow{p}\Vert
(\pm \pi ,\pm \pi )$. It is convenient to measure momentum from the fermi
point in the nodal direction and to parametrize the fermion dispersion by
two velocities: \ one, $v_{F}$, of the order of the underlying fermi
velocity describing motion perpendicular to the direction in which the gap
varies, and one, $v_{\Delta }$, describing the opening of the gap and of
order $\Delta _{0}/p_{F}$, obtaining 
\begin{equation}
E_{p}=\sqrt{(v_{F}p_{1})^{2}+(v_{\Delta }p_{2})^{2}}  \label{Edirac}
\end{equation}

We take the fermions to be normal ordered in the basis which diagonalizes $%
H_{D}$ , so the contribution of the negative energy (filled Dirac sea)
states is subsumed in the phase Lagrangian density $\mathcal{L}_{\phi }$
which we write as 
\begin{equation}
\mathcal{L}_{\phi }=\frac{1}{2}\left( \partial _{i}\phi +2ieA_{i}\right)
\ast \rho _{s0}^{ij}(r-r^{\prime })\ast \left( \partial _{j}\phi
-2ieA_{j}\right)  \label{Lphi}
\end{equation}
Here $i$ is a Cartesian direction, $\mathbf{A}$ is the vector potential and
we have allowed for non-locality in space so the $\ast $ represents
convolution.

The quantity $\rho _{s0}^{ij}$ is a diagonal matrix with dimension of $%
energy/length^{2}$. \ Its components $\ \rho _{s0}^{xx}(r)=\rho
_{s0}^{yy}(r)\equiv \rho _{s0}(r)$ (we assume tetragonal symmetry) are
related, in the absence of quasiparticle excitations, to the conventionally
defined superfluid stiffness $\rho _{s}$ (measurable, e.g. from the London
penetration depth) by 
\begin{equation}
\rho _{s}(T=0,H=0)=\int d^{2}r\rho _{S0}(r)  \label{rhos}
\end{equation}
Systems near a Mott transition are characterized by a low density of mobile
charges, and we therefore expect that $\rho _{S0}(r)$ has a length
dependence set by this low density.

The term $\mathcal{L}_{mix}$ gives the coupling of the phase fluctuations to
the nodal quasiparticles; it may be written 
\begin{equation}
\mathcal{L}_{mix}=\sum_{\alpha ,\sigma ,p,q}\left( \frac{1}{2}\partial _{\mu
}\phi (r)\\
+ieA_{\mu }(r)\right) \cdot e^{iq\cdot r}Z_{p}^{e}\overrightarrow{v}%
_{F}\psi _{p+q/2\alpha \sigma }^{+}\psi _{p-q/2,\alpha \sigma }  \label{Lmix}
\end{equation}
Here $\alpha =1...4$ labels the four nodes of the d-wave state and $Z^{e}$
is a phenomenological constant which we will show below may be thought of as
the charge of a superconducting quasiparticle. It may depend on position
relative to the node and on the proximity to the Mott transition and will be
discussed in more detail below. It has been stated in the literature that
one generically has $Z^{e}=1$; but this is now known not to be correct. Eq. $%
\mathcal{L}_{mix}$ is the long wavelength limit of a more general action
involving also 'pairbreaking' terms such as $\psi ^{+}\psi ^{+}$ with
coefficients of order $q$.

\section{Physical content}

The low energy, long wavelength theory is described by four parameters: $%
\rho _{S0}$, $v_{F}$, $v_{\Delta }$ and the quasiparticle charge
renormalization $Z^{e}$.. To see how these parameters may be determined
experimentally, we integrate out the fermions in the presence of static,
slowly varying superflow field and vector potential, which enter via the
gauge invariant combination 
\begin{equation}
\overrightarrow{Q}=\left( \overrightarrow{\nabla }\phi (r,t)-2ie%
\overrightarrow{A}\right)  \label{Qdef}
\end{equation}
and are taken to be slowly varying on the scales set by $\rho _{S}$ and the
fermions. We obtain for the two dimensional free energy density 
\begin{equation}
F_{static}(\overrightarrow{Q})=\frac{1}{2}\rho _{S}^{0}Q^{2}\\-
2T\sum_{\alpha
}\int dEN(E)\ln \left[ 1+\exp [-(E+\frac{1}{2}Z_{p}^{e}\overrightarrow{Q}%
\cdot \overrightarrow{v}_{a})/T\right]  \label{Fstatic}
\end{equation}
where the 2 is for spin, the sum $(\alpha )$ is over the four nodes of the
Dirac spectrum and we have introduced the single-node single-spin density of
states per unit area 
\begin{equation}
N(E)=\int \frac{d^{2}p}{\left( 2\pi \right) ^{2}}\delta (E-E_{p})=\frac{E}{%
2\pi v_{F}v_{\Delta }}  \label{dos}
\end{equation}

The specific heat may be obtained by differentiating Eq. \ref{Fstatic} twice
with respect to $T$ \ and is 
\begin{equation}
\frac{C}{T}=\frac{T}{4\pi v_{F}v_{\Delta }}\sum_{\alpha }\int_{0}^{\infty }dx%
\frac{x(x+\frac{Z_{p}^{e}\overrightarrow{Q}\cdot \overrightarrow{v}_{a}}{2T}%
)^{2}}{\cosh ^{2}\left[ \frac{x+\frac{Z_{p}^{e}\overrightarrow{Q}\cdot 
\overrightarrow{v}_{a}}{2T}}{2}\right] }  \label{C/T}
\end{equation}
The integral may easily be evaluated numerically for given $Q,T$. Analytical
results exist in the limits $Q/T\rightarrow 0$ and $Q/T\rightarrow \infty $.
The zero-field specific heat coefficient (per unit area) $C(B=0,T)$ is 
\begin{equation}
\frac{C(B=0)}{T}=\frac{18\zeta (3)T}{\pi v_{F}v_{\Delta }}  \label{C(B=0)/T}
\end{equation}
while in the high field low-T limit we obtain (after symmetrization) 
\begin{equation}
\frac{C(Z^{e}vQ>>T)}{T}=\sum_{\alpha =1...4}\frac{\pi Z^{e}}{%
12v_{F}v_{\Delta }}\left| \overrightarrow{Q}\cdot \overrightarrow{v}%
_{a}\right|  \label{highq}
\end{equation}
Averaging Eq \ref{highq} over the $Q$-distribution characteristic of a
vortex stateleads to the 'Volovik' prediction \cite{Volovik} of a $B^{1/2}$
field dependence of the specific heat if $B_{c2}>>B>>B_{c1}$. In principle
the result depends on the nature of the vortex state and on the relative
angle between the lattice vectors characterizing the vortex lattice (if any)
and \ the directions corresponding to the gap nodes in the superconducting
state. We have evaluated the averages and find that the dependence is weak: 
\begin{equation}
\frac{C(B>\Phi _{0}v_{F}^{2}/T^{2})}{T}=\frac{\pi Z^{e}}{3v_{\Delta }}\left( 
\frac{B}{\Phi _{0}}\right) ^{1/2}A  \label{Chigh-field}
\end{equation}
where $A=\pi /2$ for a square vortex lattice with nodal direction aligned
with the vortex lattice vector and $A=1.52$ for $45$ degrees misalignment,
similarly for a triangular vortex lattice $A=1.7$ with $5\%$ variations as
the angle is varied. Some details of the calculation are presented in the
Appendix.

Similarly the differential penetration depth is given by differentiating $F$
twice with respect to $Q$ . For an arbitrary current distribution $\rho _{S}$
is a tensor: 
\begin{equation}
\rho _{S}^{ab}=\rho _{S0}\delta _{ab}-\sum_{\alpha }\frac{TZ^{e2}v_{\alpha
}^{a}v_{\alpha }^{b}}{4\pi v_{F}v_{\Delta }}\int_{0}^{\infty }dx\frac{x}{%
\cosh ^{2}\left[ x+\frac{Z^{e}\overrightarrow{Q}\cdot \overrightarrow{v}_{a}%
}{4T}\right] }  \label{rhosgen}
\end{equation}
where $a,b$ are specific Cartesian directions and $v^{a}$ is the component
of $v_{F}$ in the $a$ direction. Eq. (\ref{rhos(B)}) describes among other
things the nonlinear Meissner effect \cite{nonlinmeiss}: note however the
importance of the charge renormalization factor $Z^{e}$.

In the weak field limit, $\rho _{S}^{ab}=\rho _{S}\delta _{ab}$ with 
\begin{equation}
\rho _{S}(T)=\rho _{S0}-\frac{\ln (2)Z^{e2}v_{F}}{2\pi v_{\Delta }}T=\rho
_{S0}-\frac{\ln (2)Z^{e2}v_{F}^{2}}{36\zeta (3)}\frac{C(B=0)}{T}
\label{rhos(T)}
\end{equation}

The factor $Z^{e}$ is essentially the Landau parameter introduced in
previous work \cite{Landaurefs}. Comparison of Eqs. \ref{C/T} and \ref
{rhos(T)} shows why $Z^{e}$ is more appropriately interpreted as the
quasiparticle charge renormalization. The usual f-sum-rule
(Ferrel-Glover-Tinkham) arguments imply that the change, with temperature,
in the condensate fraction is balanced by an increase in the 'normal'
conductivity due to quasiparticles. Now the quasiparticle conductivity is
determined by the number of carriers (which follows from the specific heat,
which essentially counts excitations) and their velocity, (which may be
determined from photoemission). Any remaining discrepancy with the observed $%
d\rho _{S}/dT$ must then be due to their charge, i.e. to the factor $Z^{e}$.
\ .

At large $Q$ and low $T$ we find the current-dependence of the superfluid
stiffness to be 
\begin{equation}
\rho _{S}^{ab}(Q,T=0)=\rho _{S0}\delta _{ab}-\frac{Z^{e3}v_{\alpha
}^{a}v_{\alpha }^{b}}{16\pi v_{F}v_{\Delta }}\sum_{\alpha }\left| 
\overrightarrow{Q}\cdot \overrightarrow{v}_{a}\right|  \label{rhos(Q)}
\end{equation}

Calculations similar to those for the specific heat yield, for a vortex
lattice with square or triangular symmetry, $\rho _{S}^{ab}=\rho _{S}\delta
_{ab}$ with

\begin{equation}
\rho _{S}(B,T=0)=\rho _{S0}-A\frac{Z^{e3}v_{F}^{2}}{4\pi v_{\Delta }}\sqrt{%
\frac{B}{\Phi _{0}}}  \label{rhos(B)}
\end{equation}

\section{Experimental values}

\subsection{Overview}

The important parameters of the theory, $\rho _{S}$, $Z^{e},v_{F},v_{\Delta
} $ may be determined from experiment. Of these, the crucial parameter is $%
Z^{e}$. Unfortunately, the present situation is unclear because different
determinations do not agree; also most measurements determine only
combinations of the fundamental quantities, so that uncertainties in one
propagate into uncertainties in another. In the following sub-sections
section we discuss the available data for each of the three parameters, and
then in a concluding subsection summarize the results and outstanding
questions.

\subsection{$\protect\rho _{s}$}

The $T=0$ superfluid stiffness has been reasonably well determined by muon
spin rotation experiments \cite{Uemura89,Sonier00} and decreases as the Mott
insulator is approached. The decrease is apparently roughly proportional to
hole doping. We regard this result as well established and we do not discuss
it further.

\subsection{v$_{F}$}

Angle-resolved photoemission measurements yield $v_{F}$ \cite{Damascelli01}%
;. At present the generally accepted value for high-T$_{c}$ materials \
(both optimal and underdoped) along the zone diagonal and in the
superconducting state is\cite{Schabel98,Johnson01} 
\begin{equation}
v_{F}=1.8[eV-A]  \label{vf}
\end{equation}
The velocity apparently increases slightly as doping is decreased. There is
general agreement concerning the value and doping independence of the
velocity (note that even undoped materials exhibit zone diagonal velocities
of approximately this magnitude). We regard this parameter as having been
reasonably reliably established.

\subsection{v$_{\Delta }$}

The parameter $v_{\Delta }$ may be obtained in three ways: from
photoemission, from zero-field specific heat, and from thermal conductivity.
Each method is subject to uncertainties, as outlined below.

Photoemission investigations of the form of the superconducting gap near the
nodes reveal a broadened structure, with a nonvanishing density of states in
a small arc around the zone diagonal \cite{ARPESUD}. This could be an
intrinsic effect, indicating a non-d-wave form of the gap function, or it
could be due to pairbreaking or other sample and surface imperfections. \
However, evidence that the gapless arcs have a non-intrinsic origin is
provided by penetration depth and thermal conductivity measurements
discussed below, so we take this view here. An estimate of $v_{\Delta }$
from photoemission may be obtained by combining the gap maximum value $%
\Delta _{0}$, the standard $\cos (2\theta )$ d-wave form and the arc length
from the zone diagonal fermi point to the gap maximum point, which is
roughly $\pi /\sqrt{2}b$ with $b$ the lattice constant, leading to 
\begin{equation}
v_{\Delta }=\frac{2\sqrt{2}b\Delta _{0}}{\pi }  \label{vdelta}
\end{equation}
Estimates for the gap maximum range from $30-40meV$ in optimal $YBCO$ (with
the large values in the direction parallel to the chains and the smaller in
the direction perpendicular \cite{Lu01} to $40meV$ in $BSCCO$ \cite{Mesot99}
leading to 
\begin{eqnarray}
v_{\Delta } &=&0.13[eV-A]\;(BSCCO)  \label{vdelvalBSCCO} \\
v_{\Delta } &=&0.09-0.12[eV-A]\;(YBCO)  \label{vdelvalYBCO}
\end{eqnarray}
Available photoemission evidence \cite{Mesot99} suggests that $\Delta _{0}$
and therefore $v_{\Delta }$ if anything increase with decreasing doping;
suggesting (if we interpret the maximal gap observed in the $(\pi ,0)$
direction as superconducting gap) that $v_{\Delta }$ increases with
decreasing doping. These estimates rely on the assumption that everywhere in
the zone the observed gap\ has a superconducting origin. While this
assumption has been used by many workers, and appears to be supported by the
good argeement between the simple d-wave form and the data of \cite{Mesot99}%
, diffferent interpretations exist in which the gap in underdoped materials
has a non-superconducting origin \cite{DDW,Varma99}.

Eq. (\ref{C(B=0)/T}) shows that measurements of the low temperature specific
heat yield the product $v_{F}v_{\Delta }$. Because we regard the value of $%
v_{F}$ as reliable, these measurements yield a value for $v_{\Delta }$. In
optimally doped $YBCO,$ specific heat data \cite{Wang01} yield (in present
notations \cite{conventions}) 
\begin{equation}
v_{F}v_{\Delta }=0.06\left[ eV-A\right] ^{2}  \label{vfvdeltawang}
\end{equation}
or 
\begin{equation}
v_{\Delta }=0.033\left[ eV-A\right]  \label{vdeltawang}
\end{equation}

This value is far outside the range of $v_{\Delta }$ suggested by
photoemission. The authors of Ref. \cite{Wang01} suggest that the
discrepancy occurs because there are additional contributions to the \
measured low-field specific heat (for example from chain states) which
should not be included in the comparison between the model and data. This
idea is consistent with recent microwave conductivity measurements \cite
{Harris01} which find evidence for a large density of gapless excitations
associated with the chains. An alternative possibility is that the gap
function does not have the simple $\cos (2\theta )$ form often assumed, but
instead is less strongly angle dependent near the nodes, so that $v_{\Delta
} $ is not well estimated from the maximum gap value. Reliable measurements
of the low temperature specific heat for $BSCCO$ or underdoped $%
YBCO$ are not available.

Thermal conductivity measurements yield values for $v_{F}/v_{\Delta }$ \cite
{Taillefer,Chiao00,Tailleferunpub}. These results rely upon a theoretical
'universal limit' expression for the low temperature limit of a transport
coefficient \cite{Durst00}, and upon the assumption that this low
temperature limit \ has been experimentally accessed. \ Measurements\ \cite
{Taillefer} yield $v_{F}/v_{\Delta }=19$ for optimally doped BSCCO. For YBCO
a strong doping dependence is found. As doping is decreased the ratio drops
from about 19 for a presumably slightly overdoped $YBCO_{6.993}$ \cite
{Tailleferunpub}sample to 14 for a putatively optimally doped $YBCO_{6.95}$ 
\cite{Chiao00} to 8 for the $60$ phase $YBCO_{6.73}$. \cite{Tailleferunpub}.
These estimates suggest that $v_{\Delta }$ rapidly increases with
underdoping. 
\begin{eqnarray}
v_{\Delta } &=&0.095\;(BSCCO,overdopedYBCO)  \label{vdelthermbscco} \\
v_{\Delta } &=&0.13\;(optimally\,doped\,YBCO)  \label{vdeltermybco} \\
v_{\Delta } &=&0.2(underdoped\,YBCO)
\end{eqnarray}
These data are roughly consistent with the $v_{\Delta }$ inferred from the
gap maximum found in photoemission; however one should bear in mind that the
increase in $v_{\Delta }$ found as doping is decreased corresponds to a
decrease in the value of the 'universal limit' thermal conductivity. This
could arise from an inhomogeneous sample (in which not all of the material
is superconducting) or possibly from novel physics (not included in the
basic action studied here) suppressing the ability of the quasiparticles in
a doped Mott insulator to carry heat.

\subsection{Z$^{e}$}

The crucial quantity $Z^{e}$ appears in combination with $v_{F},v_{\Delta }$
and so values are subject to uncertainties, particularly in the value of $%
v_{\Delta }$.

The temperature dependence of the penetration depth yields the combination $%
\frac{Z_{e}^{2}v_{F}}{v_{\Delta }}$. In YBCO certainly and 
in other high-$T_c$ materials, probably,  the temperature
dependence of the penetration depth in the direction transverse to the
chains (if any) is only weakly material-dependent, and is linear at low $T$
with the slope given by \cite{Bonn96,BSCCO}

\begin{eqnarray}
\frac{d\rho _{S}}{d\left( k_{B}T\right) } &\approx &0.7\hspace{0.11in}%
(YBCO6.6,6.9)  \label{drhosdtYBCO} \\
&\approx &0.9\hspace{0.11in}(BSCCO,optimal)  \label{drhosdtbscco}
\end{eqnarray}
Note that this linearity is inconsistent with the presence of the ''gapless
arcs'' \cite{ARPESUD} in the electronic spectrum. From Eq. \ref{rhos(T)} we
then obtain

\begin{equation}
\frac{Z_{e}^{2}v_{F}}{v_{\Delta }}=6-8  \label{Z2v}
\end{equation}
The ability to determine $Z^{e}$ by combining penetration depth data with
values for $v_{F}$ and $v_{\Delta }$ (obtained for example from thermal
conductivity data) \ was noted by Chiao, Taillefer and co-workers \cite
{Taillefer}; the values obtained \ from the thermal conductivity data
discussed above then yield 
\begin{eqnarray}
Z^{e} &=&0.7\;(Optimal\,BSCCO;\,Overdoped\,YBCO)\,  \label{zebscco} \\
Z^{e} &=&0.8\;(Opt\,imally\ doped\,YBCO)  \label{zeoptybco} \\
Z^{e} &=&1\;(60K\,\ YBCO)  \label{zeudybco}
\end{eqnarray}

The magnetic field dependence of the specific heat yields $\frac{v_{\Delta }%
}{Z_{e}A}$ where $A$ is a constant (discussed above) relating to the current
distribution in the vortex lattice. In optimally doped $YBCO,$ high-field
specific heat data \cite{Wang01} yield (in present notations \cite
{conventions}) 
\begin{equation}
\frac{v_{\Delta }}{Z_{e}A}=0.09\left[ eV-A\right]  \label{vdeloverza}
\end{equation}
Use of our estimate $A\approx 1.7$ \cite{conventions} and the range quoted
above for $v_{\Delta }$ yields 
\begin{equation}
0.6<Z^{e}<1
\end{equation}

Recent microwave conductivity measurements \cite{Harris01} reveal an
additional difficulty with the quantitative extraction of $Z^{e}$ in $YBCO$:
in this material the deviations from tetragonal symmetry are found to lead
to a strong ($\sim 50\%$) variation in the plane conductivity (which can be
separated from the chain conductivity) between electric field parallel to
the chain direction and antiparallel to it. \ This anisotropy has not been
taken into account in our analysis.

\subsection{Summary}

In summary, at present the experimental status of the parameter values
characterizing the superconducting state is not completely satisfactory. The
specific heat results for optimal $YBCO$ suggest rather smaller $v_{\Delta }$
values than are found by other measurements, and photoemission and some
tunnelling data suggest that $v_{\Delta }$ decreases as doping is reduced,
while other measurements including thermal conductivity suggest that it
increases. The available data suggest however that the crucial parameter $%
Z^{e}$ is of order unity and is only weakly dependent on doping.
Particularly compelling in this regard is the observed weak doping
dependence of $d\rho _{S}/dT$, combined with the doping independence of $%
v_{F},$ and the indications that $v_{\Delta }$ increases with decreasing
doping. These indications suggest that $Z^{e}$ is of order unity and if
anything increases as doping is decreased. Data contradicting this
conclusion exist. Further experimental information would be very helpful.

\section{Range of applicability of low energy action}

\subsection{Overview}

The results presented above constitute the leading temperature and $Q$
dependence about the $T=0,Q=0$ limit because they are nonanalytic in the
standard expansion parameters $(T/E_{0})^{2},(v_{F}Q/E_{0})^{2}$ where $%
E_{0} $ is a 'microscopic' energy scale (for example the BCS gap amplitude $%
\Delta _{0}$). We expect the expansion ceases to hold when the correction
terms are of the order of the leading terms and in particular when the
corrections to $\rho _{S0}$ are of the order of $\rho _{S0}$.

One source of correction terms are terms of the order of $Q^{4}$ in the
phase part of the action. The usual expectation from study of quantum
critical points is that these become important when 
\begin{eqnarray}
Q &\sim &Q^{\phi }=\left( \rho _{S0}/E_{\phi }\right) ^{1/2}/\xi _{0}
\label{Qphi} \\
T^{\phi } &\sim &\rho _{S0}  \label{Tphi}
\end{eqnarray}
where $E_{\phi },\xi _{0}$ are 'microscopic' energy and length scales which
do not vanish as the Mott phase or other critical point is approached.

Another correction occurs when the fermionic terms become of the order of
the leading terms, i.e. when 
\begin{equation}
T\sim T^{\ast }=\rho _{S0}\left( \frac{v_{\Delta }}{Z_{e}^{2}v_{F}}\right)
\label{Tlim}
\end{equation}
or 
\begin{equation}
Q\sim Q^{\ast }=\frac{2\rho _{S0}}{v_{F}}\left( \frac{v_{\Delta }}{%
Z_{e}^{3}v_{F}}\right)  \label{Qlim}
\end{equation}

Roughly, if the fermionic terms determine the limits of validity of the low $%
T,Q$ expansion, then the physics of the nonsuperconducting state is
dominated by electrons and is expected to be more or less a conventional
metal, whereas if the phase terms set the limits then fermions are
irrelevant at the superconducting-non-superconducting critical point and the
physics is presumably bosonic.

In general, the limit of validity of the low $T$ low $B$ expansion signals
the destruction of the superconducting state. We shall discuss the
superconducting non-superconducting transition on the assumption that the
physics is strictly two dimensional. While this is a reasonable
approximation for high-T$_{c}$ materials, it is important to bear in mind
that ultimately a crossover to three dimensional critical behavior will
occur and that the parameter controlling the crossover is the inverse of the
square of the logarithm of the superfluid stiffness anisotropy $1/\ln
^{2}(\rho _{S\Vert }/\rho _{S\bot })$ which is not extremely small in
practice, so although the two dimensional arguments provide reasonable
estimates of the energy scales controlling $T_{c}$ and (as discussed below) $%
H_{c2}$, a quantitative application requires some caution.

\subsection{Thermal fluctuations, zero field}

In a two dimensional material the thermally driven zero-field
superconducting-non-superconducting transition is a Kosterlitz-Thouless
vortex unbinding transition. It occurs at a $T_{KT}$ satisfying $%
T_{KT}=2\rho _{S}(T_{KT})/\pi $. Because thermal effects can only decrease $%
\rho _{S}$ from its $T=0$ value, the scale $T^{\phi }$ defined in Eq. \ref
{Tphi} is an upper bound for this transition temperature.

Two kinds of thermal effects occur: fluctuations of the superconducting
phase, and quasiparticle excitations. In the absence of a high density of
quasiparticle excitations, longitudinal \'{(}''spin-wave'') phase
fluctuations involve unscreened charge fluctuations and are therefore
strongly suppressed by the Coulomb interaction. In the absence of a high
density of quasiparticles the only important excitations are
vortex-antivortex pairs, whose energetics are governed by the scale $\rho
_{S0}$. (In this regard we worry that the numerical studies of Ref. \cite
{Carlson00} are not quantitatively relevant to superconductors because these
studies were based on the classical $XY$ model, so a large contribution from
'spin-wave' longitudinal excitations is apparently included, whereas one
would expect these to be strongly suppressed in an electronic system in
which coulomb forces were important). Quasiparticle excitations will reduce $%
\rho _{S}(T)$ from its $T=0$ value. If $\frac{v_{\Delta }}{Z_{e}^{2}v_{F}}<1$
then the limit $T^{\ast }$ set by quasiparticle effects is more stringent.
The physics of this limit \ (which seems to be favored by the data) is
simply that thermal quasiparticle excitations reduce $\rho _{S}$ so that the
Kosterlitz-Thouless transition occurs at a temperature lower by a factor of $%
F$ than one would guess from $\rho _{S0}$.

\subsection{Field driven effects: low T}

Application of a magnetic field $B>B_{c1}$ produces vortices in the
superconducting order parameter. A superconducting vortex consists of a
''core'' and a ''far'' region. In the far region superconducting excitation
spectrum is only weakly perturbed and there is a circulating supercurrent of
a magnitude $j_{S}\rho _{S}/r$. As the core region is approached the
supercurrent magnitude exhibits a maximum and then decreases and the
quasiparticle excitation spectrum fundamentally changes. These two effects
are distinct and define two core sizes, $\xi _{curr}$ at which $dj_{S}/dr=0$
and $\xi _{exc}$ at which the excitation spectrum changes. In a conventional
superconductor $\xi _{exc}\approx $ $\xi _{curr}=v_{F}/\Delta $. In the high 
$T_{c}$ context, substantial attention has focused on $\xi _{exc}$ (which is
apparently very short \cite{Renner98}) and on the possibility that the
change of the excitation spectrum is not simply a collapse of the
superconducting gap (as in conventional materials) but instead involves the
appearance of a new form of long range order, for example antiferromagnetism
or staggered flux \cite{Arovas97,Han00,Lee01,Kishine01}. Here we wish to
focus on $\xi _{curr}$ which in a lightly doped Mott insulator may be much
greater than $\xi _{exc}$. Writing $j_{S}^{a}=\delta F/\delta Q^{a}$ and
using Eq \ref{Fstatic} leads to 
\begin{equation}
j_{S}^{a}=\rho _{s0}Q^{a}-\frac{Z_{e}^{3}}{8\pi v_{F}v_{\Delta }}%
\sum_{\alpha }v_{\alpha }^{a}\left( \overrightarrow{v}_{\alpha }\cdot 
\overrightarrow{Q}\right) ^{2}\Theta \left( \overrightarrow{v}_{\alpha
}\cdot \overrightarrow{Q}\right)  \label{js}
\end{equation}

Taking $Q$ to be parallel to a gap node and of magnitude $1/r$ we find that
the current is maximal at $\xi _{curr}=\frac{v_{F}^{2}Z^{3}}{4\pi \rho
_{S0}v_{\Delta }}$ \textit{provided} that $\xi _{curr}$ is greater than the
scale over which $\rho _{S0}$ varies. Eq \ref{js} shows again the importance
of $\ $\ the doping dependence of $Z^{e}$. If (as available data seem to
suggest), $Z^{e}$ remains constant and $\rho _{S0}\sim x$ then $\xi
_{curr}\sim x^{-1}$, whereas gauge-theory based models \cite{Ioffe88}
(including, we believe, those discussed in \cite{Lee96,Wen98}) lead to $%
Z^{e}\sim x$ so that $\xi _{curr}$ is controlled by the scale dependence of $%
\rho _{S},$ implying $\xi _{curr}\sim x^{-1/2}$. Typical numbers for
optimally doped $YBCO$ are $\rho _{S0}\sim 10meV$, $v_{F}/v_{\Delta }\sim 15$%
, implying $\xi _{curr}[A]\approx 100Z^{3}$, roughly consistent with muon
spin rotation data \cite{Sonier00}, although of course uncertainties in $Z$
lead to large uncertainties in the numerical estimates.

The magnitude of $\xi _{curr}$ is important because $H_{c2}$ is essentially
the field at which the vortex cores overlap, and for the resistive
transition it is natural to use the 'current' definition of the vortex core
size. Essentially this argument was given by Lee and Wen \cite{Lee96} who
were among the first to emphasize the importance, in the high-T$_{c}$
context, of the scale over which the supercurrent varied and (on the
assumption that $Z^{e}=1$) concluded that $H_{c2}\sim x^{2}$. \ Future
papers will examine in more detail the assumption that $\xi _{curr}$ is the
correct measure of the core size to use in estimating $H_{c2}$, but the
plausibility of this claim may be seen for example from Eq. (\ref{rhos(B)})
which shows that when field becomes large enough to suppress $\rho _{S}$ by
a factor of order unity the intervortex spacing is of the order of $\xi
_{curr}$.

\section{Conclusions}

We have presented and compared to data a general theory of low energy
properties of a d-wave superconductor. The theory has four parameters: the $%
T=0$, $H=0$ superfluid stiffness $\rho _{S0}$, the velocities $v_{F}$ and $%
v_{\Delta }$ describing the Dirac spectrum of d-wave quasiparticles, and a
quantity $Z^{e}$ which expresses the coupling between quasiparticles and
phase fluctuations and which we argued should be interpreted as the charge
of the nodal quasiparticle. The behavior of these quantities contains
information about the physics of the approach to the Mott transition,
because different theoretical treatments of doped Mott insulators predict
(or assume) quite different variations of these parameters with doping. The
behavior of these quantities controls many aspects of the physics: in
particular, the size (as defined from the supercurrent distribution) of
superconducting vortices.

Two widely discussed theoretical approaches are the Brinkman-Rice-dynamical
mean field theory \cite{Georges96} and the slave boson gauge theory \cite
{Ioffe88} The essential ingredient of the Brinkman-Rice theory is a
self-energy $\Sigma $ with a strong frequency dependence and a negligible
momentum dependence. This leads to a Mott transition driven by a divergent $%
\partial \Sigma /\partial \omega $ implying $Z^{e}$ independent of $x$ and $%
v_{F}\sim x$. The latter prediction is in apparent contradiction to
photoemission data. The essential assumption of the gauge theory approach is
that current is carried by a small density of holes doped into a spin liquid
environment. The fermionic excitations of the superconducting state are
combinations of hole and spin-liquid states and the low density of holes
leads to a small charge $Z^{e}\sim x$. In other words, the quasiparticles
become more neutral as the Mott insulating phase is approached. One may
think of this as a precursor of the 'nodal liquid' phase discussed in \cite
{Nodal}. This idea also appears to be inconsistent with the available data,
although, as emphasized in Section III the available data are not entirely
consistent. Further, and perhaps most important, \ complete information is
not yet available for underdoped (especially strongly underdoped) materials.
We urge the experimental community to settle the issue of the data
consistency, in order to finally establish the relevance of the
Brinkman-Rice and gauge theory approaches to the physics of high-T$_{c}$.

Our understanding of the presently available data favors the hypothesis that 
$Z^{e}$ and $v_{F}$ remain constant as $\rho _{S}\rightarrow 0$,. This
result would appear to rule out both the Brinkman-Rice and gauge theory
descriptions of the Mott physics of high $T_{c}$ materials, and it is
therefore interesting to understand the origin of the discrepancy. One
common feature of the two approaches is that in them the Mott physics
affects all of the Fermi surface in the same manner, so the reduction in
charge stiffness is described by a reduction in velocity or in quasiparticle
charge. If neither of these effects occurs, then the reduction in charge
stiffness must be driven by a reduction in 'effective fermi surface area'.
In other words, it seems likely that in high $T_{c}$ materials the crucial
missing ingredient is a large, doping dependent variation of the parameters
around the Fermi surface so that all superfluid properties arise from
condensation of fermions in a narrow and doping dependent range around the
nodal direction.

Consider for example Eq. (\ref{rhos(Q)}) which describes the reduction of
the $\rho _{S}$ due to depairing of the nodal quasiparticles by a non-zero
superflow. A phase gradient of magnitude $Q$ depairs electrons in an angular
range $\delta \theta \sim Z^{e}v_{F}Q/v_{\Delta }$. In an underdoped
material it seems that $Z^{e}$ remains of order unity while $\rho _{S}$
becomes very small. Eq. (\ref{rhos(Q)}) then implies that exciting
quasiparticles in a range $\delta \theta \ll \pi /2$ will reduce $\rho _{S}$
to zero, i.e. that all or most of the supercurrent is carried by electrons
small patches, of angular size $\delta \theta \sim \rho _{S}v_{\Delta
}/Z^{e}v_{F}^{2}$ centered on the nodal points. Within this picture an
interesting question is the behavior of $Z$ for angles $\theta >\delta
\theta $. Because $\rho _{S}$ cannot become negative, the quasiparticles
must in some manner decouple from the superfluid fluctuations (i.e $Z^{e}$
must become small in these regions).

There is to our knowledge no microscopic theory of the narrow patch
situation described above which is consistent with all data. One possibility
is a commensurate long range order which opens a large, doping dependent gap
around the antinodal points $(\pi ,0)$ which kills most of the Fermi surface
leaving only hole pockets around the diagonals which then acquire a small
amplitude superconducting gap. One example of this phenomenon would be the
'd-density wave' state. Another would be some form of antiferromagnetic or
'stripe' order. Two crucial consequences of such physics are a broken
symmetry (which should be detectable in various spectroscopies) and a small $%
v_{\Delta }$ (determined by the observed $T_{c}$). We think that the
available data do not favor this proposal. Crucially, the thermal
conductivity measurements suggest that $v_{\Delta }$ increases when doping
is decreased. An alternative which is at least qualitatively consistent with
the data is preformed (d-wave) pairs which are made mostly from the
electrons near $(\pi ,0)$ regions and which do not contribute to any
transport. For example, \cite{Geshkenbein96} proposed a theory in which a
large mass in the $(\pi ,0)$ region prevented the gap maximum regions from
contributing to transport. We see here that an alternative is a small $Z^{e}$%
. Unfortunately there is no controlled microscopic theory which yield this
physics, although uncontrolled but interesting extrapolations of scaling
equations have been argued to lead to this physics \cite{Furukawa98}.

To summarize: elucidation of the experimental support (or lack thereof) for
the 'patch picture' and (assuming it is relevant) clarification of its
theoretical basis are two important challenges for the future.

\textit{Acknowledgements:} We thank D. Bonn, P. A. Lee, M. Chiao and L.
Taillefer for very helpful discussions and L. Taillefer for sharing
unpublished data. We acknowledge support from NSF-DMR-00081075.

\section{ Appendix: Current distribution in vortex lattice: H$%
_{c1}<<H<<H_{c2}$}

\subsection{Formalism}

In the limit $H_{c2}>>H>>H_{c1}$ we have, for the supercurrent distribution, 
\begin{equation}
\overrightarrow{j}_{s}(r)=\rho _{S}\overrightarrow{Q}(r)  \label{jbasic}
\end{equation}
with $\rho _{S}$ the superfluid stiffness and the gradient of the
superconducting phase field given by 
\begin{equation}
\overrightarrow{Q}(r)=\sum_{i}\frac{\widehat{z}\times \left( \overrightarrow{%
r}-\overrightarrow{r}_{i}\right) }{\left( \overrightarrow{r}-\overrightarrow{%
r}_{i}\right) ^{2}}  \label{qvort}
\end{equation}
The quantity appearing in the expression for the field-induced specific heat
for a vortex lattice oriented at angle $\theta $ to the gap node direction
is 
\begin{equation}
C(\theta )=\frac{1}{A_{V}}\int dxdy\left| Q_{x}\cos (\theta )+Q_{y}\sin
(\theta )\right|  \label{C}
\end{equation}
where the integral is over the unit cell of the vortex lattice and $A_{V}$
is the area of this cell. The result has dimension of $length^{-1}$. It is
convenient to measure lengths in units of the inter-vortex spacing $a$ and
to normalize the result to the square root of the vortex density $%
n_{V}=B/\Phi _{0}$. Thus we write 
\begin{equation}
C\left( \theta \right) =n_{v}c(\theta )  \label{c}
\end{equation}
and compute $c(\theta )$ for square and triangular vortex lattices.

\subsection{Numerical evaluation, square vortex lattice}

We consider a square vortex lattice of lattice constant $a$, so the vortices
sit at positions $na\widehat{x}+ma\widehat{y}.$ $n_{v}=a^{-2}$. Eq \ref
{qvort} gives 
\begin{eqnarray}
Q_{x} &=&n_{V}^{1/2}\sum_{n,m}\frac{(m-y/a)}{\left( n-x/a\right) ^{2}+\left(
m-y/a\right) ^{2}}  \label{qxsquare} \\
Q_{y} &=&-Q_{x}(y,x)  \label{qysquare}
\end{eqnarray}
Consider $Q_{x}$. The sum over $y$ is most conveniently evaluated in Fourier
space by writing $\sum_{m}\rightarrow \int du\rho (u)$ with $\rho
(u)=a^{-1}\sum_{m}\delta (y-ma)=\sum_{k}e^{i2\pi ky}$ Substitution gives 
\begin{equation}
Q_{x}(x,y)=\frac{\pi }{2}n_{V}^{1/2}\sum_{n}\frac{\sin (2\pi y/a)}{\sinh
^{2}(\pi \left( n-x/a)\right) +\sin ^{2}\left( \pi y/a\right) }  \label{ysum}
\end{equation}
The sum on $n$ is rapidly convergent and may easily be evaluated
numerically. \ 

We wish to evaluate Eq. \ref{C} by integrating over the region $-a/2<x,y<a/2$%
. This is most conveniently evaluated numerically by inscribing a circle in
the unit cell, performing the integral over the circle in polar coordinates
(to eliminate the divergence at $r\rightarrow 0$) and then integrating over
the remaining regions. This latter integral is over the region $-a/2<y<a/2$; 
$a/2>\left| x\right| >\sqrt{\frac{a^{2}}{4}-y^{2}}.$ We have performed this
integral numerically using Mathematica; results are shown in the Table below.

\subsection{Triangular lattice}

Lattice vectors 
\begin{eqnarray}
\mathbf{v}_{1} &=&a\widehat{x}  \label{basis1} \\
\mathbf{v}_{2} &=&a\left( \frac{-1}{2}\widehat{x}\mathbf{+}\frac{\sqrt{3}}{2}%
\widehat{y}\right)  \label{basis2}
\end{eqnarray}
A general lattice vector is then $n\mathbf{v}_{1}+m\mathbf{v}_{2}$. \ The
unit cell is a hexagon with area $3\sqrt{3}a^{2}/8$.

Eq. \ref{qvort} then gives, for the component of $\mathbf{j}$ perpendicular
to $\mathbf{v}_{1}$ 
\begin{equation}
\mathbf{Q}_{y}=-\sum_{m,n}\frac{n-\frac{1}{2}m-x}{\left( n-\frac{1}{2}%
m-x\right) ^{2}+\left( \frac{\sqrt{3}}{2}ma-y\right) ^{2}}  \label{jxtriang}
\end{equation}

The sum over $n$ may again be performed--it is just the previous result with 
$y\rightarrow x+m/2$ and $n-x/a\rightarrow \frac{\sqrt{3}}{2}ma-y$ so that 
\begin{equation}
Q_{y}(x,y)=\frac{\pi }{2a}\sum_{m}\frac{\sin (m\pi +2\pi x/a)}{\sinh
^{2}(\pi \left( \frac{\sqrt{3}}{2}m-y/a)\right) +\sin ^{2}\left( m\pi /2+\pi
x/a\right) }  \label{qytriang}
\end{equation}
$Q_{x}$ is obtained by computing the component perpendicular to a different
basic lattice vector and then combining appropriately.

For the Volovik effect we require $\int_{hexagon}dxdy\left| \mathbf{v\cdot j}%
(x,y)\right| .$ We find this is very well approximated by the integral over
the inscribed circle. Results are shown in the Table.

\begin{tabular}{|c|c|c|}
\hline
Angle & c$_{square}$ & c$_{triangle}$ \\ \hline
0 & $1.5708...$ & 1.74 \\ \hline
$\pi /8$ & 1.55 & 1.72 \\ \hline
$\pi /4$ & 1.52 & 1.71 \\ \hline
\end{tabular}

\textit{Table caption: }Coefficient $c$ defined in Eq \ref{c} for square and
triangular vortex lattice as function of angle $\theta $ between lattice
vector of vortex lattice and nodal direction of d-wave superconducting order
parameter.


\begin{thebibliography}{99}
\bibitem{Orenstein00}  For a review, see J. Orenstein and A. J. Millis,
Science \textbf{288} 468-74 (2000).

\bibitem{Imada98}  M. Imada, A. Fujimori and Y. Tokura, Rev. Mod. Phys. 
\textbf{70}, 1039-1263 (1998).

\bibitem{Lee96}  P. A. Lee and X.-G. Wen Phys. Rev. Lett. \textbf{78}, 4111
(1997).

\bibitem{Tseui98}  C. C. Tseui and J. R. Kirtley, Rev Mod Phys \textbf{72 }%
969-1016 (1998).

\bibitem{dinunderdoped}  Underdoped high-$T_{c}$ materials have not been
tested for d-wave superconductivity as thoroughly as optimally doped
materials. The linear temperature dependence of the penetration depth in $%
60K $ $YBCO$ (\cite{Bonn96})) and the generally d-wave-like form of the gap
observed via photoemission in underdoped BSCCO (see Shen and Dessau, ref
below or \cite{Mesot99}), along with the argument of continuity from
optimally doped materials provide perhaps the strongest evidence.

\bibitem{Shen95}  Z. X. Shen and D. S. Dessau \ Phys. Rep. \textbf{253}, 1
(1995).

\bibitem{Taillefer}  M. Chiao, R. W. Hill, C. Lupien, L. Taillefer, P.
Lambert, R. Gagnon,and P. Fournier, Phys. Rev. \textbf{B62 }3554-8 (2000)

\bibitem{Uemura89}  Y. J. Uemura et. al., Phys. Rev. Lett. \textbf{62}
2317-20 (1989).

\bibitem{Kotliar}  B.G. Kotliar and J. Liu, Phys. Rev \textbf{B38}, 5142
(1988); Y. Suzumura et. al. J. Phys. Soc Jpn \textbf{57}2768-72 (1988).

\bibitem{Emery95}  V. J. Emery and S. A. Kivelson, Nature \textbf{374}, 434
(1995).

\bibitem{Corson98}  J. Corson et al., Nature \textbf{398}, 221 (1999).

\bibitem{Volovik}  G. Volovik, J.E.T.P. Lett., \textbf{58} 469 (1993).

\bibitem{nonlinmeiss}  S-K. Yip and J. Sauls, Phys. Rev. Lett. \textbf{69}
2264-7 (1992).

\bibitem{Landaurefs}  A. I. Larkin, Sov. Phys. JETP; A. J. Leggett, Phys.
Rev.; A. J. Millis, S. \ M. Girvin, L. B. Ioffe and A. I. Larkin, J. Phys.
Chem. Solids \textbf{59}, 1742-5 (1998).

\bibitem{Wen98}  X. G. Wen and P. A. Lee, Phys. Rev. Lett. \textbf{80}, 2193
(1998)

\bibitem{Sonier00}  J. E Sonier, Jess H Brewer and Robert F Kiefl, Rev. Mod.
Phys. \textbf{72}, 769 (2000).

\bibitem{Damascelli01}  For a recent review see e.g. A Damascelli, D. H. Lu
and Z. X. Shen, J Electron Spectrosc. Relat. Phenom. \textbf{117-8} pp165-87
(2001).

\bibitem{Schabel98}  M. C. Schabel, C-H Park, A Matsuura, Z-X Shen, D.A.
Bonn, Ruixing Liang, and W. N. Hardy, Phys. Rev. \textbf{B57 }6090 (1998).

\bibitem{Johnson01}  Johnson, P.D.; Valla, T.; Fedorov, A.V.; Yusof, Z.;
Wells, B.O.; Li, Q.; Moodenbaugh, A.R.; Gu, G.D.; Koshizuka, N.; Kendziora,
C.; Sha Jian; Hinks, D.G., Phys. Rev. Lett. \textbf{87 }177007 (2001).

\bibitem{ARPESUD}  M. Randeria and J. C. Campuzano, in Proceedings of the
International School of Physics ''Enrico Fermi,'' Varenna 1997,
(North-Holland, New York); H. Ding et. al. J. Phys. Chem. Sol. \textbf{59 }%
1888-91 (cond-mat/97121000).

\bibitem{Lu01}  D. H. Lu et. al., Phys. Rev. Lett. \textbf{86}, 437003
(2001).

\bibitem{Mesot99}  J. Mesot et. al., Phys. Rev. Lett. \textbf{83} 840-3
(1999).

\bibitem{DDW}  S. Chakravarty, R. B. Laughlin, D. K. Morr and C. Nayak,
Phys. Rev. \textbf{B63} 094503/1-10 (2000).

\bibitem{Varma99}  C. M. Varma, Phys. Rev. Lett. \textbf{83 }3538-41 (1999).

\bibitem{Wang01}  Y. Wang, B. Revaz, A. Erb and A. Junod, Phys. Rev. \textbf{%
B63} 094508 (2001).

\bibitem{conventions}  To relate the present conventions to those of Wang 
\textit{(op. cit.)} it is simplest to note that Wang et al give results for $%
v_{F}v_{\Delta }$ and for $v_{\Delta }/a$ and that the relation of their
parameter $a$ to our parameter $A$ may be determined from the ratio of the
zero field and high field specific heats. We find $a=\pi ^{2}A/8.$ The value 
$a=0.7$ found by Wang et. al is thus rather smaller than the $A=1.74$
calculated here, leading to a larger value of $Z^{e}$.

\bibitem{Harris01}  R. Harris et. al. Phys. Rev. \textbf{B64 }\ 064509-17
(2001).

\bibitem{Durst00}  A. Durst and P. A. Lee, Phys. Rev. \textbf{B62 }1270
(2000).

\bibitem{Chiao00}  May Chiao, R W Hill, Christian Lupien, Phys. Rev. Lett. 
\textbf{82}, 2943 (2000).

\bibitem{Tailleferunpub}  L. Taillefer, private communication.

\bibitem{Bonn96}  D. Bonn et. al. Czech Journal of Physics . \textbf{46},
S6, 3195 (1996).

\bibitem{BSCCO}  The data on non-YBCO materials is less extensive. $La_{2-x}
Sr_xCuO_{4+\delta}$ is perhaps the best studied but effects of disorder
and magnetic and charge ordering complicate the picture. Data for
one and three-layer $Hg$ materials are presented and discussed in
C. Panagopoulos, J. R. Cooper and T. Xiang, Phys. Rev. {\bf B57}
13422-5 (1998).

\bibitem{Carlson00}  E W Carlson, S A Kivelson and V J Emery, Phys. Rev.
Lett. \textbf{83}, 612 (1999).

\bibitem{Renner98}  Ch Renner, B Revaz, K Kadowaki, Phys. Rev. Lett. \textbf{%
80}, 3606 (1998)

\bibitem{Arovas97}  D. P. Arovas, A. J. Berlinsky, C. Kallin, and S.-C.
Zhang, Phys. Rev. Lett. \textbf{79}, 2871-2874 (1997)

\bibitem{Han00}  Jung Hoon Han, Dung-Hai Lee, Phys. Rev. Lett. \textbf{85},
1100-3 (2000).

\bibitem{Lee01}  P. A. Lee and X. G. Wen, Phys. Rev. \textbf{B63 }224517
(2000).

\bibitem{Kishine01}  Jun-ichiro Kishine, Patrick A Lee and Xiao-Gang Wen,
Phys. Rev. Lett.\textbf{\ 86}, 5365 (2001).

\bibitem{Georges96}  A. Georges, G. Kotliar, W. Krauth, and M. J. Rozenberg,
Rev. Mod. Phys. \textbf{68}, 13-125 (1996).

\bibitem{Ioffe88}  L. B. Ioffe and A. I. Larkin, Phys. Rev. \textbf{B39},
8988-8999 (1989)

\bibitem{Nodal}  L. Balents, M. P. A. Fisher and C. Nayak, Phys. Rev. 
\textbf{B61 }6307-19 (2000).

\bibitem{Geshkenbein96}  V. B. Geshkenbein, L. B. Ioffe, and A. I. Larkin,
Phys. Rev. B \textbf{55}, 3173-3180 (1997).

\bibitem{Furukawa98}  .Nobuo Furukawa, T M Rice and Manfred Salmhofer, Phys.
Rev. Lett.\textbf{\ 81}, 3195 (1998).
\end{thebibliography}
\end{document}